# The Making of the Chandra X-ray Observatory: the Project Scientist's Perspective

**Martin C. Weisskopf** *

*NASA/Marshall Space Flight Center

**We review the history of the development of the Chandra X-ray Observatory from our personal perspective. This review is necessarily biased and limited by space since it attempts to cover a time span approaching 5 decades.**

It is sobering for me to realize that there are scientists who are working with data from this truly great observatory who were not even born when the foundation for what is now the Chandra X-ray Observatory was laid. Thus, it may surprise many to know that the beginning was succinctly and accurately outlined in a research proposal that Riccardo Giacconi and colleagues wrote in 1963, a mere 9 months after he and his team's discovery of the first extra-solar X-ray source Scorpius X-1. As important, the data from this rocket flight also indicated the presence of the "diffuse X-ray background". Figure 1 illustrates the showpiece of this insightful proposal. It shows an approximately 1-meter diameter, 10-meter focal length, grazing-incidence X-ray telescope. The telescope was of sufficient area and angular resolution to determine the nature of the unresolved X-ray background. We all owe Riccardo an enormous debt of gratitude for his insight, leadership, and, in my case (and I suspect for many others) inspiration.

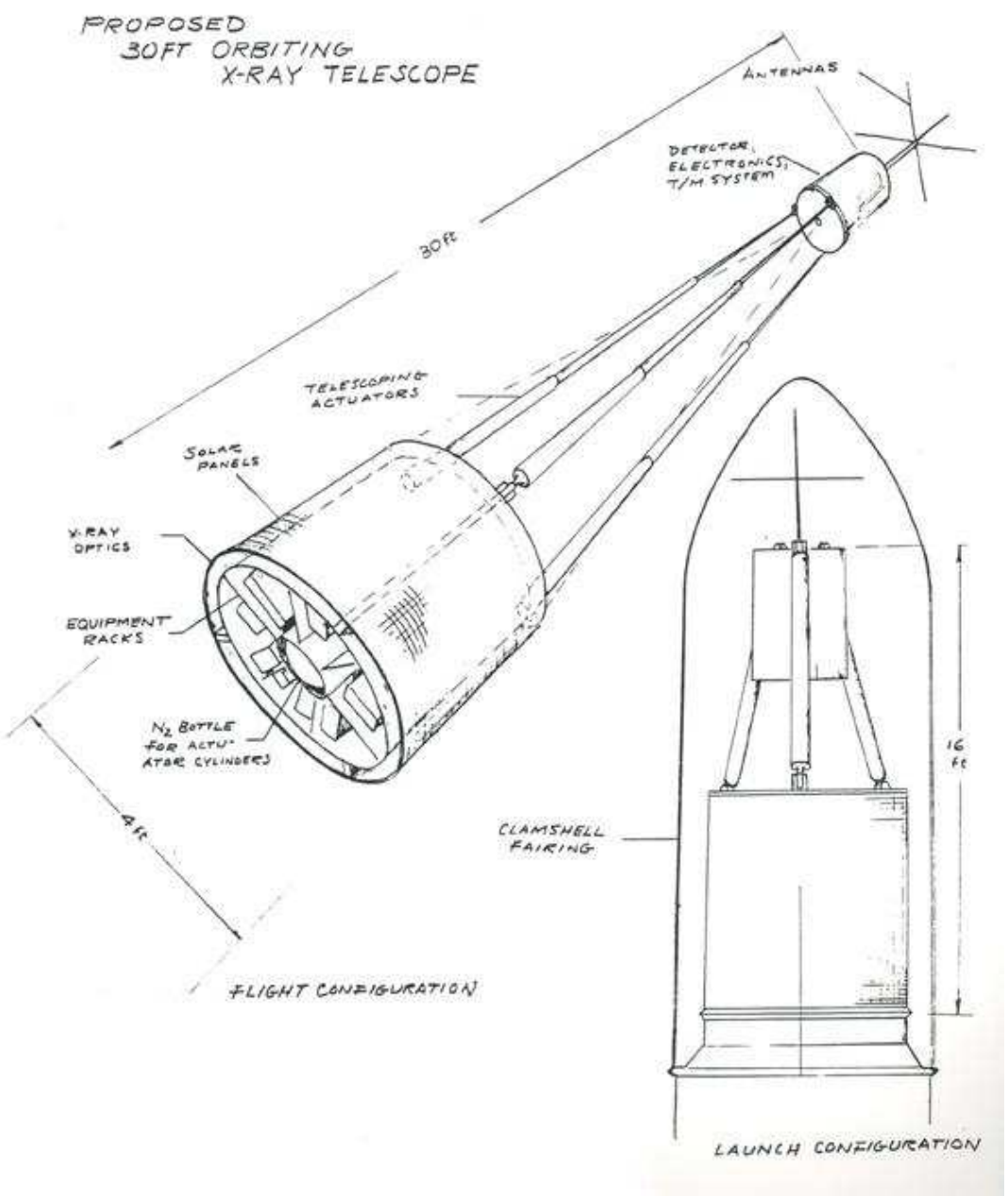

**Fig. 1.** Drawing from the 1963 Proposal.

The resemblance between the early conceptual design shown in Figure 1 and Chandra is no accident and is of importance in considering the way we (the scientific community) design our missions. Chandra was based only on achieving its scientific requirements, principally to be able to resolve the faint background sources. Chandra was not built on flying "what we can do". Neither in 1963 nor indeed in 1976 — when Riccardo and his Co-Principal Investigator Harvey Tananbaum submitted their unsolicited proposal "For the Study of the 1.2 Meter X-ray Telescope National Observatory" — did one actually know how to build the sub-arc-second telescope required to meet the scientific objectives. I feel it is important for one to know that these objectives were never compromised during the entire 23-year development, measured from the submission of this proposal to the launch in 1999. Nor did they lose their relevance during this time. This is in contrast to many (but not all e.g. the Wide Field X-ray Telescope) missions, which suffer from what I term "cost-credibility paranoia" — wherein one can only convince others of the cost reliability of the mission if one has essentially already built it. In too many cases I feel this approach has forced one to compromise scientific objectives and to adopt a "we will build it, you will use it" approach to science. In these cases too often are the scientific requirements adjusted to be compatible with the existing technology, as opposed to driving the technology. The approach isn't terrible, as missions such as the Rossi Timing X-ray Explorer have had, despite outdated technology, a high measure of success. Nevertheless, Chandra is an outstanding example of the power of the science-driven approach.

The proposal that Riccardo and Harvey submitted in 1976 drew attention at NASA headquarters, which then initiated a competition amongst the NASA Centers to establish where such a mission might best be accomplished. In reaching a decision, NASA considered such factors as expertise, experience, manpower availability, facilities, etc. The Marshall Space Flight Center (MSFC) teamed with the Smithsonian Astrophysical Observatory (SAO), the Jet Propulsion Laboratory (JPL) teamed with the California Institute of Technology (CIT); and the Goddard Space flight Center (GSFC) teamed with GSFC scientists in vying for the mission. I joined NASA in 1977, after MSFC was assigned the responsibility for the mission. I did this with the understanding that Project Science was to be more than a single person and that the local project science team would be further supplemented by the group at SAO which became known as the Mission Support Team (MST). There can be no question that the outstanding success of Chandra is due in no small part to these arrangements.

The first Science Working Group of what was then called the Advanced X-ray Astrophysics Facility was chaired by Riccardo. I served as Vice-Chairman. Other members and their affiliations were: A. Opp (NASA HQ; ex officio), E. Boldt (GSFC), S. Bowyer (University of California Berkeley), G. Clark (Massachusetts Institute of Technology), A. Davidsen (John Hopkins University), G. Garmire (California Institute of Technology), W. Kraushaar (University of Wisconsin-Madison), R. Novick (Columbia University) , S. Shulman (Naval Re-

**Reserved for Publication Footnotes**

search laboratory), H. Tananbaum (SAO), A. Walker (Stanford), K. Pounds (Leicester University), and J. Trümper (Max Planck Institut für Extraterrestriche Physik).

The peculiar name (AXAF) of the Project was the inspiration of the NASA Associate Administrator at the time. He did not want to use the word "telescope" in describing a future program because Congress had recently approved a telescope — what is now known as the Hubble Space telescope (HST).

1979 saw the launch of the Einstein Observatory and its subsequent astounding impact on astronomy and astrophysics — namely that all categories of objects from comets to quasars were X-ray sources and that study of their X-ray emission provides critical insights into emission mechanisms, evolution scenarios, etc. This in turn led to the report of the Astronomy Survey Committee "Astronomy and Astrophysics for the 1980's" to recommend "An Advanced X-ray Astrophysics Facility (AXAF)" as the number one priority for large, spaced-based missions. The recommendation was more profound than one might imagine, as there was only one X-ray astronomer (George Clark) on the committee. Despite this superlative recommendation, it would take almost two decades before Chandra launched. Although there were numerous reasons for this slow advance, and I am over-simplifying, Chandra appeared to need to wait its turn while HST was being developed. We see this pattern reemerging as the International X-ray Observatory (IXO) appears to need to await the completion of the James Webb Space Telescope (JWST). A famous Nobel Prize winner has referred to the apparent delays in some programs caused by financial overruns and technical problems of other missions as "the punishment of the innocent", in a recent white paper to the current Decadal Survey.

1983 saw the release of the Announcement of Opportunity for the first set of instrumentation — the "first set" as AXAF was envisioned as being serviceable at that time. Experiment selection took place in 1985 and a new science working group was formed which I chaired and whose members and are pictured in Figure 2.

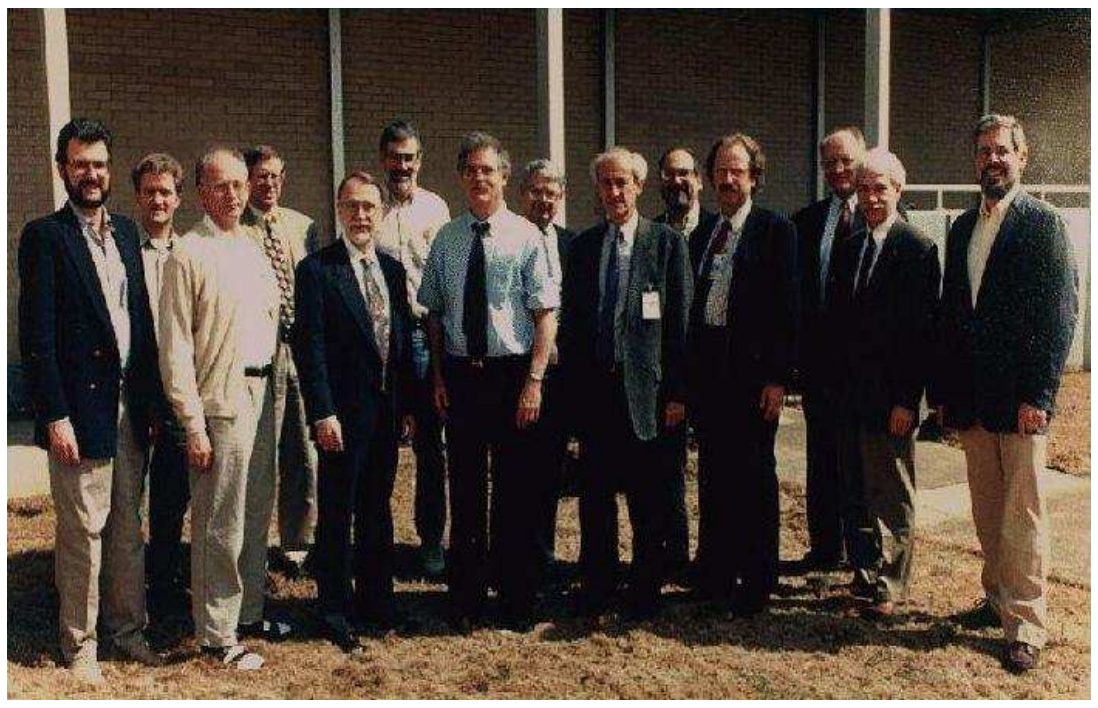

**Fig. 2.** The second Science Working Group. Left to right: A. Wilson (Interdisciplinary Scientist (IDS)), A. Fabian (IDS), J. Linsky (IDS), H. Tananbaum (head of the MST and ultimately Director of the Chandra X-ray Center), A. Bunner (ex officio), S. Holt (XRS PI), M. Weisskopf (Project Scientist), R. Giacconi (IDS), A. Brinkman (LETG PI), S. Murray (HRC PI), G. Garmire (ACIS PI), L. van Speybroeck (Telescope Scientist), C. Canizares (HETG/FPCS PI), R. Mushotzky (IDS)

The selected instruments included the Advanced Camera for Imaging Spectroscopy (ACIS); the High Resolution Camera (HRC), the Low-Energy Transmission Grating (LETG); the High Energy Transmission Grating (HETG); the Focal-Plane Crystal Spectrometer (FPCS), which was removed during a descoping exercise in 1988; and the X-ray Spectrometer (XRS), which was moved to a mission named AXAF-S in 1992 prior to cancellation of AXAF-S by Congress in 1993.

In addition to pursuing mirror and detector technology development during the 1980's many of us also became salesmen and brochure writers, in order to gain full support for the program at NASA and in Congress. I won't speak here of the heroic role played by the Director of our Science Center, Harvey Tananbaum. There can be no question that we all owe him a huge debt of gratitude. Of course Harvey wasn't the only one who helped create this program. Others, such as Charles Pellerin and Arthur Fuchs at NASA HQ and our industrial allies, also played a significant role. It may seem inappropriate to some, but it is definitely true that accomplishment of "big money science" requires many skills. The scientific excellence of the mission is necessary but, unfortunately not sufficient. One of our most successful endeavors in this time frame was a brochure, a page of which is illustrated in Figure 3. A facing page (not shown) was designed for the intelligent layman with an inset for our colleagues in the scientific community. The cartoons (one is shown in the figure) were jokingly described as being aimed at NASA officials and Congresspersons. Nevertheless these cartoons were scientifically accurate and drew attention. The brochure not only won an artistic prize, but also made a positive impression where it needed to. During this period, Space Station was included as an ingredient in the servicing of Chandra. For a number of technical and programmatic reasons, this tie-in soon disappeared.

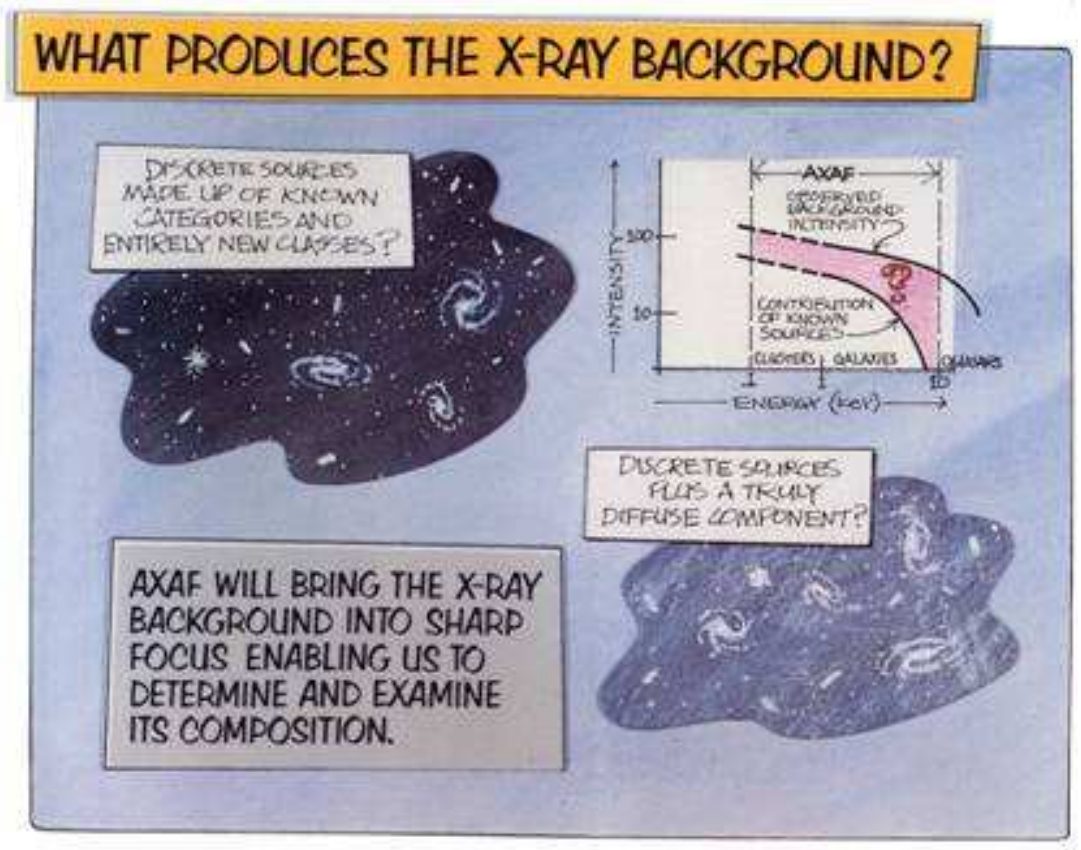

**Fig. 3.** A page from the AXAF brochure of 1985.

The late 1980's saw significant technical progress. Two versions of what we called the "Technology Mirror Assembly (TMA)", were built and X-ray tested. The assembly was comprised of a single paraboloid-hyperboloid pair. The TMA was a 2/3-scaled version of the most challenging (the innermost mirrors) AXAF mirrors. At 2/3 scale, the focal length became 6-m which allowed for testing in the existing X-ray Calibration Facility (XRCF) at MSFC. This facility had been built to calibrate and test the Einstein X-ray optics.

TMA-1 was received July 27, 1985 and the angular resolution was better than 0.5 arcseconds. However, to our chagrin, near-angle scattering due to mid-frequency errors negatively impacted the encircled energy. We had made the cardinal mistake of assuming that we needn't be concerned about errors on the mm-scale lengths that were the culprits, because we hadn't envisaged that the tools and processes we were using could introduce terms at those frequencies. Of course we changed our technical approach and removed these errors. TMA-2 was received Jan 6, 1989 and is shown in Figure 4. The performance of TMA-2 was outstanding with respect to all specifications and convinced us that the Chandra optics could be built successfully. It is interesting to know that, twenty years later, TMA-2 is still the best X-ray telescope in the world considering only angular resolution and encircled energy.

Unfortunately, our critics did not agree that we knew how to build the optics and imposed the challenge to us of manufacturing the largest set (paraboloid-hyperboloid) of AXAF optics and to demonstrate that the performance would meet requirements. In his eagerness to comply with this challenge, which would become a Congressional mandate, the Associate Administrator of NASA promised that not only would we build and test these mirrors using the optical metrology we had verified by X-ray testing the TMA, but that we would also perform the necessary X-ray tests. Moreover, he committed us to accomplish this by summer 1991. Unfortunately, the plan and funds to enhance the existing XRCF were not compatible with this schedule!

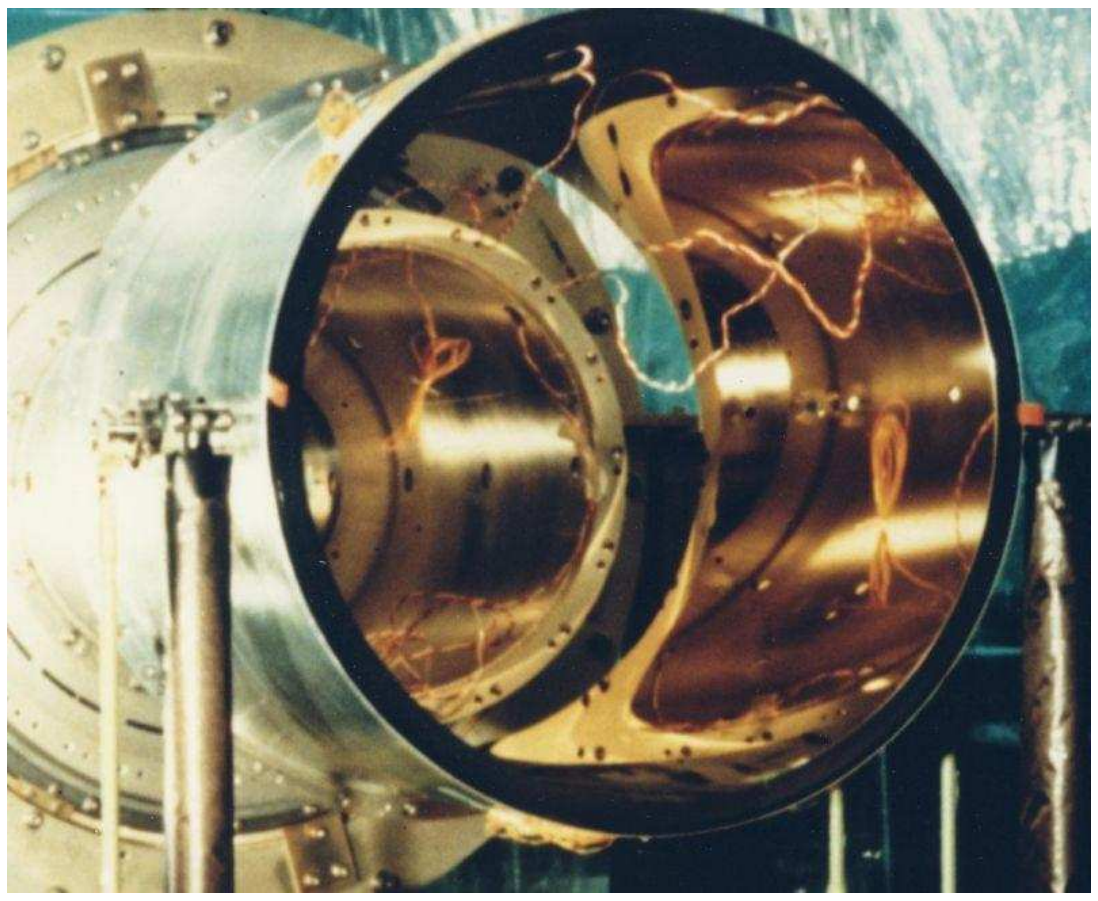

**Fig. 4.** The gold-coated TMA.

NASA always seems to work best in a crisis. The Director of MSFC took it upon himself to manage the effort, which meant meetings once a week at 7:00 a.m. in his office. Despite the early hour, the Director's intervention meant that all of the resources of the Center were available to us and without (most of) the usual bureaucratic holdups that occur when one wants to do something quickly. In my oral presentation, I showed a number of slides that documented the development of the XRCF. The interested reader may find these at http://cxc.harvard.edu/ChandraDecade/proceedings/. Figure 5 shows the completed facility. The XRCF is currently being used for visible-light, cryogenic testing of the JWST beryllium mirrors.

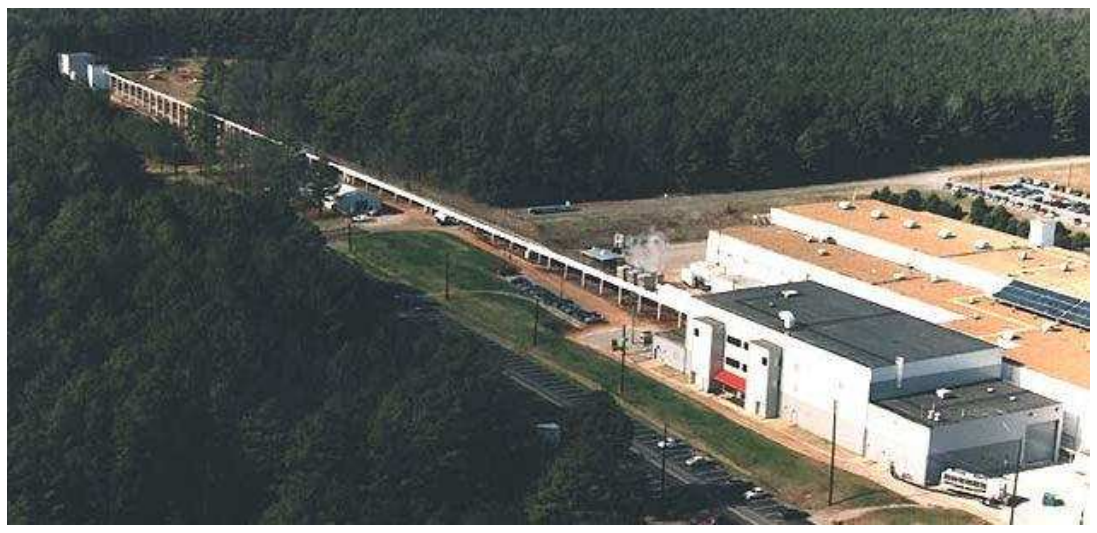

**Fig. 5.** The XRCF at MSFC. The building to the left houses the X-ray sources, filters, and monochrometers. In the building on the right is the large thermal vacuum chamber, at a distance of 525 m from the source building. The chamber accommodates the Chandra 10-m focal length and full illumination of the 1.2 meter diameter optics.

Figure 6 shows the Verification Engineering Test Article (VETA), comprised of the largest paraboloid-hyperboloid pair, but uncoated and uncut to their final length and thus not in the flight mount. Testing took place in 1991 and — after compensating for the finite distance, the size of the X-ray source, and gravitational effects — produced the outstanding result shown in Figure 7.

The Project's reward for this fantastic effort was to have the following year's budget cut, necessitating another launch delay. Despite all of the progress, both in the optics and the concurrent instrument development, the launch of Chandra had been postponed at the pace of one year per calendar year for many years. Something had to be done.

The outcome of about a year of grueling discussions and trade studies as to the details of the mission, was to abandon servicing. In order to convince the powers-that-be that servicing was not lurking in the background as a hidden variable, low-earth orbit was also abandoned. (It is interesting to contemplate the potential servicing of Chandra in the future when the NASA develops vehicles capable of traveling to the Moon and Mars. It is also sobering to realize that Chandra's first servicing would have been scheduled around the time of the Columbia accident.)

Abandoning low-earth orbit had numerous implications, some positive and some negative. The negative aspects included the sacrifice of two of the six nested telescopes (as part of the weight-reduction program necessitated by the much higher orbit) and the loss of servicing and instrument replacement. Another consequence turned out to be the establishment and subsequent loss of the AXAF-S mission, which accommodated the extremely heavy X-ray Calorimeter. This mission was canceled in 1993 by Congress, which suggested that the calorimeter be flown on a Japanese mission, albeit with poorer angular resolution than had been planed for AXAF-S. One benefit from the redesign was the switch of the telescope coating from gold to iridium. This change retained the higher-energy effective area despite the removal of the two mirrors. Another benefit was improved efficiency for observing, afforded by the orbit. This orbit reduced both the debilitating effects of occultation by the Earth and the amount of time spent in the radiation belts. The tremendous decrease in the number of occultations of the sun by the earth also greatly simplified the thermal design of the Observatory.

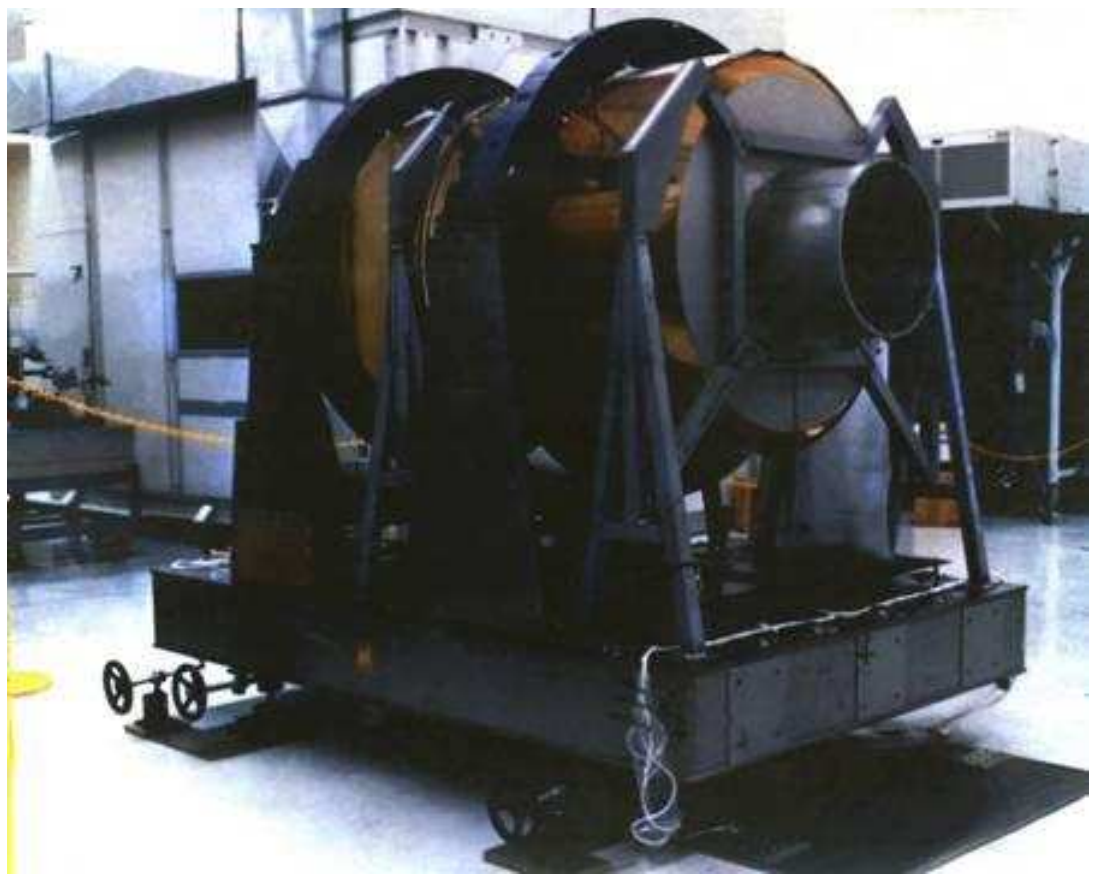

**Fig. 6.** The VETA.

The very early 1990's also saw the selection of the Chandra X-ray Center (CXC), after an open competition for an organization to serve as the interface between the Observatory and the community. This early selection would guarantee that the new organization could (and would) influence the design and development of Chandra and be in place in time to be challenged to analyze independently data taken during X-ray calibration. Since Chandra would be much more scientifically powerful, than previous missions, of course it needed to be thoroughly and precisely calibrated.

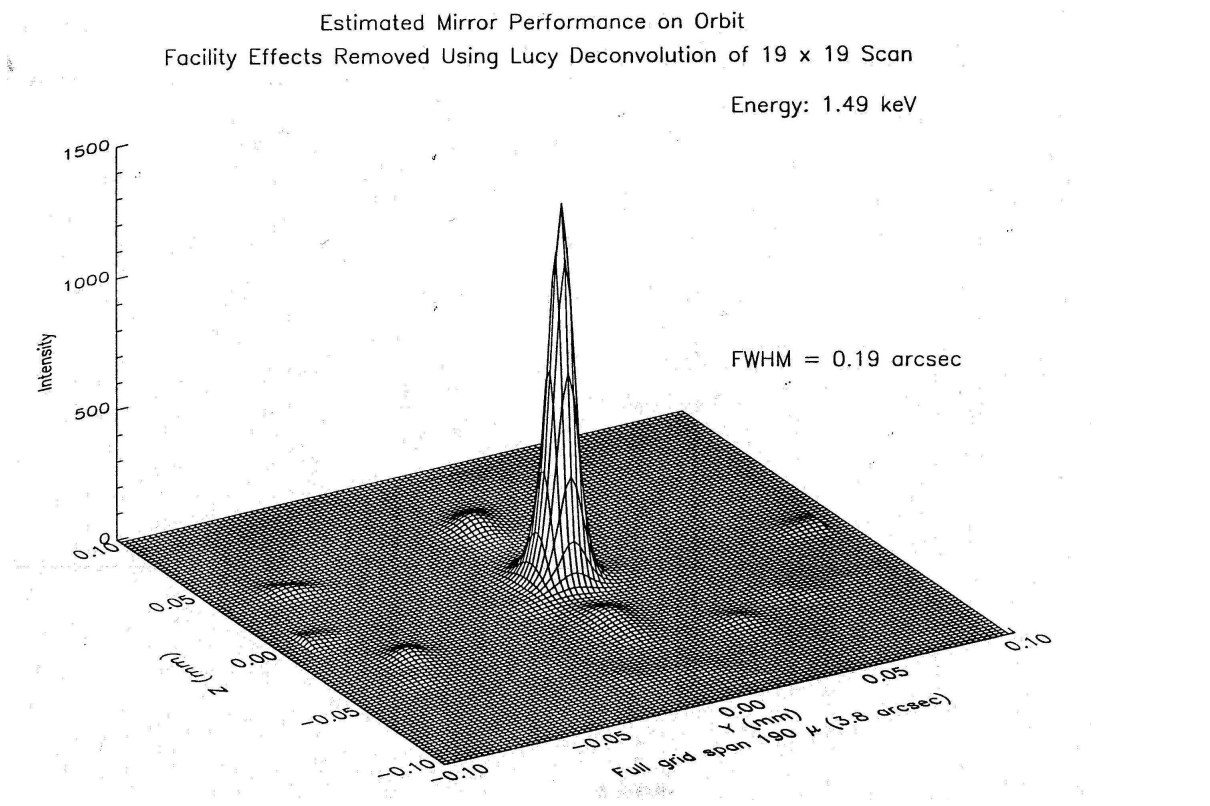

**Fig. 7.** Performance of the VETA deconvolved to remove the effects of a finite-sized source at a finite distance and showing the on-orbit performance. The FWHM is 0.19 arc-seconds. Both the FWHM and the encircled energy were well within specifications.

An important and little known development took place as part of the Chandra Program in 1992. This marked the beginning of the efforts by John Carlstrom (Cal Tech at the time) and Marshall Joy (Project Science at MSFC) to use mm-wave interferometry as a tool to measure the Sunyaev-Zeldovich (SZ) effect (Figure 8), where photons from the 3-degree microwave background are Compton-scattered by the hot X-ray-emitting electrons that pervade the intracluster medium. The scientific project of the Telescope Scientist, Leon van Speybroeck, was based on the recognition that for relaxed clusters, where the assumption of hydrostatic equilibrium was reasonable, one could simply combine X-ray measurements of the gas temperature and cluster size with SZ measurements, to determine the distance. The only difficulty, even as late as 1992, was that the SZ measurements were extremely difficult to accomplish. Thus, it was natural for Chandra and MSFC to sponsor this development.

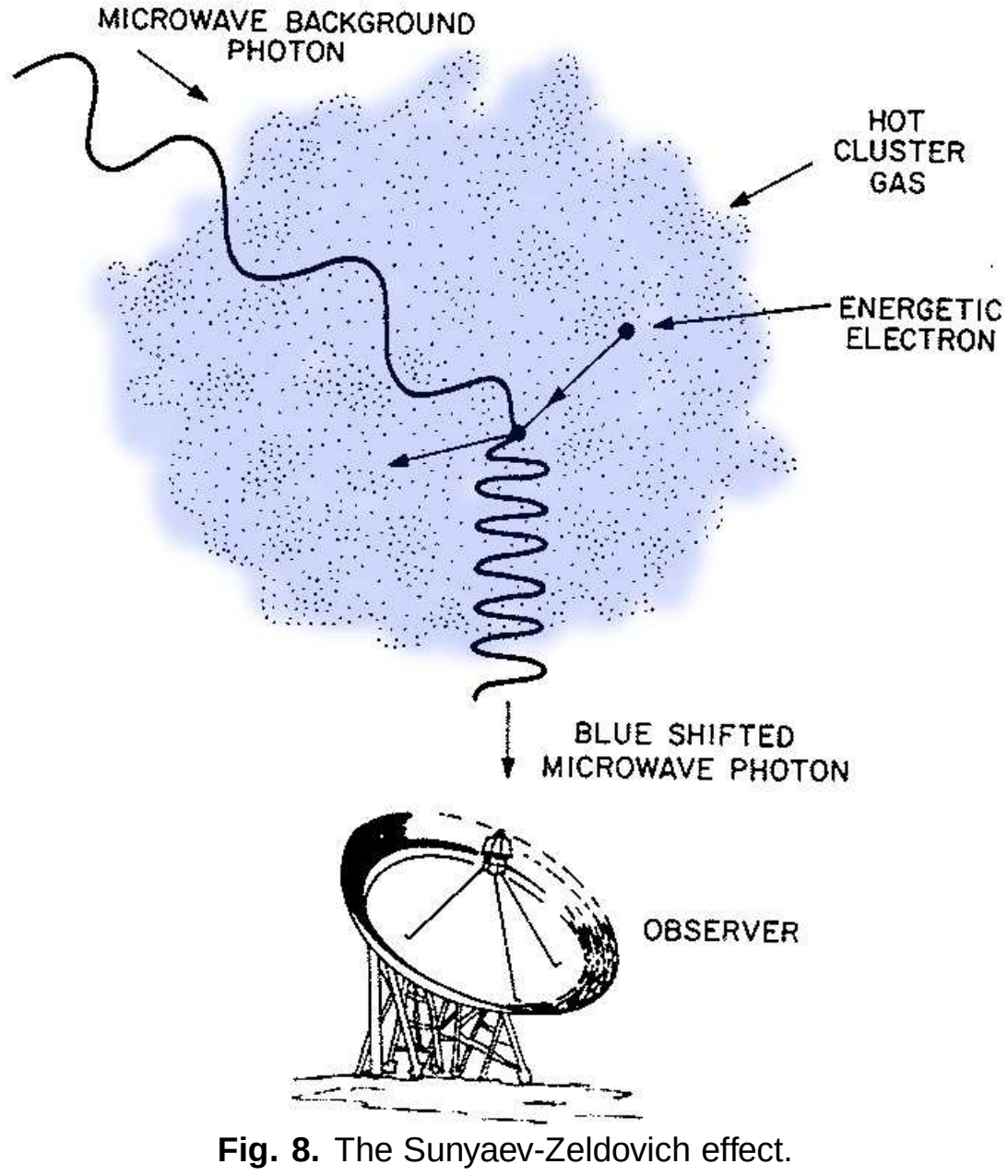

**Fig. 8.** The Sunyaev-Zeldovich effect.

With the advent of the interferometric techniques, and the subsequent development of arrays specifically designed for these observations — such as the Consolidated Array for Millimeter AstRonoMy (CARMA shown in Figure 9) and the South Pole Telescope — Chandra spawned a new scientific "industry" and was able to achieve Leon's objectives and more.

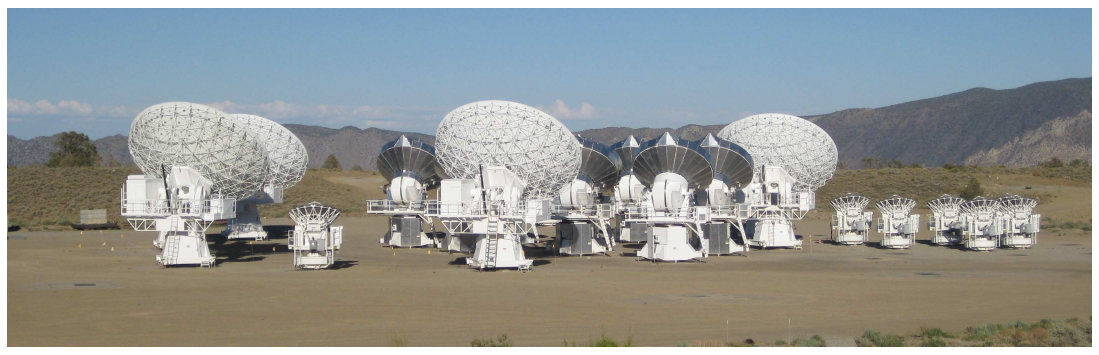

**Fig. 9.** CARMA

In the fall of 1996, the flight mirrors, fully cut, coated, aligned and mounted (known as the High Resolution Mirror Assembly or HRMA) arrived at MSFC for X-ray testing and calibration. X-ray calibration was extremely important for a number of reasons: First, the calibration activity would verify, beyond any doubt, that the optics had been built (as it turned out) to much better than their specifications. Second, and more important, the performance characteristics of the optics and the optics in conjunction with the flight instruments were calibrated to various degrees of precision as called for in a huge calibration requirements document prepared by the scientific participants. (Many have mistakenly thought that all Chandra calibration requirements were to be at the 1% level. The 1% value is mistakenly quoted out of context from the overall requirement to calibrate accurately the Observatory.)

Additional benefits of calibration included providing a data base of flight-like data, and thus an early test of the PI's and the CXC's ability to deal with flight-like data. Additional benefits were the camaraderie and friendship that developed as scientists, engineers and managers from the various different teams (MSFC Project Science (PS), SAO MST, HRC, ACIS, LETG, HETG, MSFC Project Management, MSFC Engineering, TRW, Ball Aerospace, Eastman Kodak, etc.) worked together 24/7 throughout the over-six-month period. The experience was a major contributor towards changing people's perception of each other, in particular in becoming "we" as opposed to "us and them".

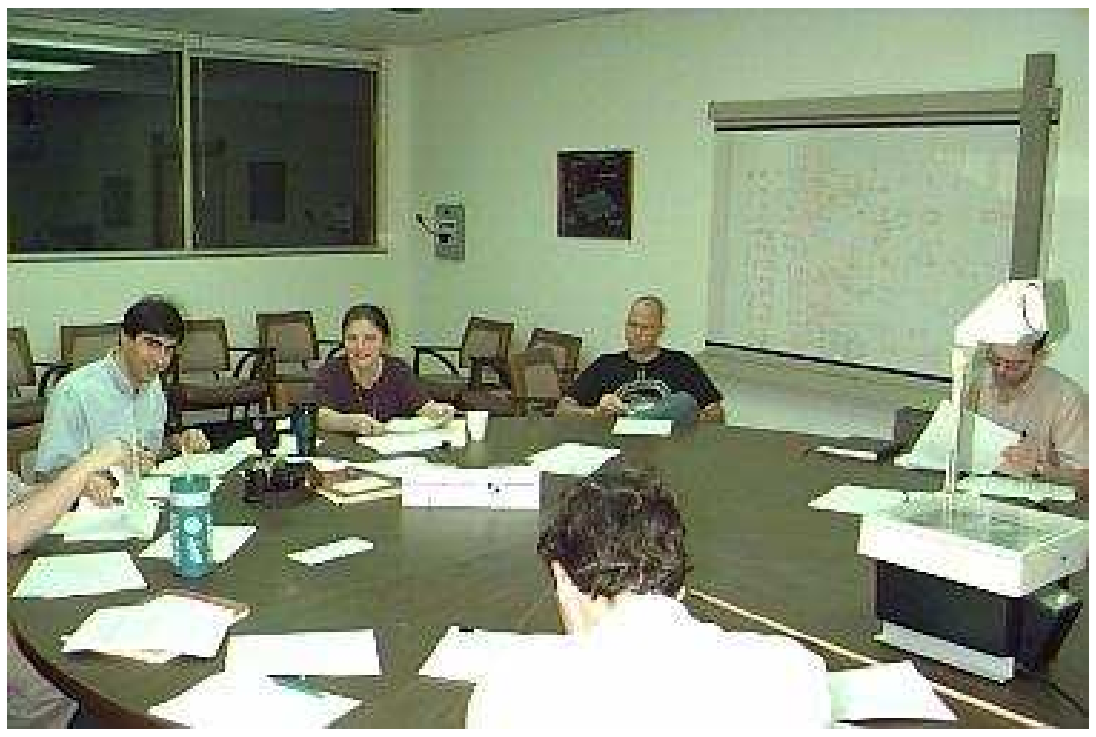

**Fig. 10.** Photograph of a typical daily meeting in the conference room at the XRCF. Clockwise: hands of S. O'Dell (PS), S. Wolk (CXC), C. Jones (CXC), R. Carlsen (TRW), D. Jerius (MST), and the back of the head of D. Dewey (HETG).

Subsequent to X-ray testing, the focal-plane detectors went to Ball Aerospace in Colorado for integration and testing into the module that houses them and provides the linear translations used both to switch instruments and to provide focus adjustment. These mechanical drives, which we worried about a lot prior to launch, have now worked routinely thousands of times throughout the mission. This Science Instrument Module (SIM) then was delivered to TRW in Cal-ifornia for integration into the spacecraft (S/C) and for system-level testing.

System-level testing, both at the SIM and S/C level, brought on a number of technical and programmatic challenges. One of most notable was the failure of the vacuum door that protected the ACIS instrument to open during thermal vacuum test. A second involved compromised printed wiring boards that required the Project to delay the launch by many months, while the appropriate electronics were removed from the S/C, replaced, retested, and re-integrated. We all owe a large debt of gratitude to Michael Hirsch at TRW for forcefully bringing this problem to everyone's attention.

In the spring of 1999 the Observatory was delivered to the the Kennedy Space Flight center for integration with the Inertial Upper Stage (IUS) and subsequent integration into the cargo bay of Columbia. The IUS is a solid-rocket, two-stage engine that would boost Chandra towards its present orbit. The combined Chandra-IUS system became the heaviest payload ever launched by a Shuttle. The activities surrounding the launch were themselves very interesting. The Commander of Columbia was Eileen Collins, the first female to hold such a position. She, along with her crew are pictured in Figure 11. Eileen drew a lot of attention and special visitors. Not only did the First Lady, Hillary Clinton, come to visit but also the singer Judy Collins wrote, composed, and then sang: "And we will fly beyond the sky. Beyond the stars, beyond the heavens. Beyond the dawn we'll carry on. Until our dreams have all come true. To those who fly - we sing to you." Perhaps as important from our perspective was the press's attention to Eileen. This attention minimized the number of times we were asked if this telescope would have problems such as HST initially experienced. Of course, we could answer that the telescope had been successfully tested and calibrated.

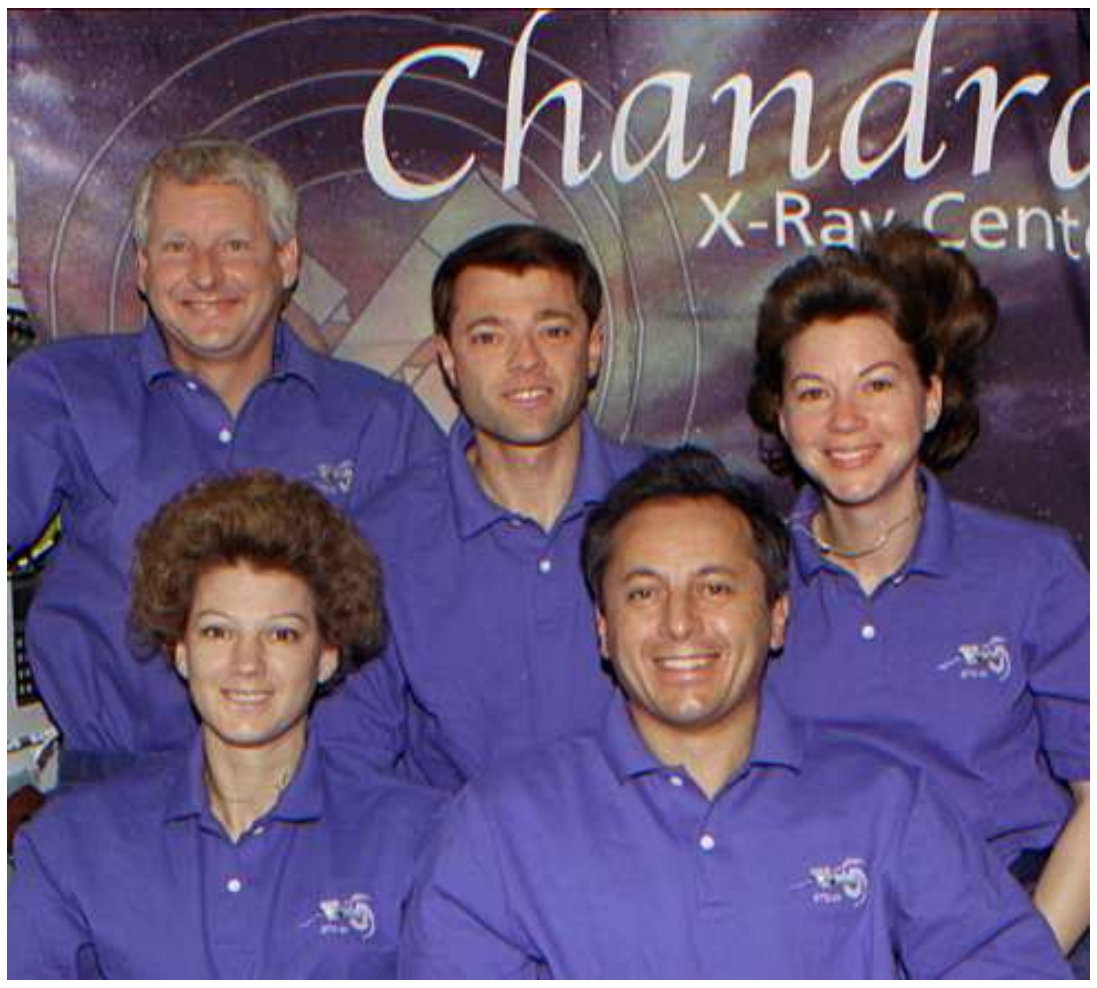

**Fig. 11.** Lower: Eileen M. Collins, Mission Commander; Michel Tognini, Mission Specialist. Upper: Steven A. Hawley, Mission Specialist; Jeffrey S. Ashby, Pilot; Catherine (Cady) G. Coleman, Mission Specialist.

Other notables at the launch included Mrs. Chandrasekhar (Figure 12), who read a poem she had composed and who noted that Prof. Chandrasekhar, after whom the Observatory is named, would not have enjoyed all the fuss. The male model Fabio also attended under the false assumption that he had been invited by the astronaut Cady Coleman. The invitation had been jokingly issued in her name by her fellow crew members.

It took three attempts to launch the Shuttle. These occurred just past midnight on the mornings of July 20, 22, and 23. The successful launch was challenging for the astronauts, due to a hydrogen leak and an electrical short which took place shortly after the rockets fired. The calmness of the astronauts through these major glitches was testimony to their courage and ability. Placement into low earth orbit was only the beginning for us of many tense days before we could be sure that the Observatory was working properly. Just eight hours into the mission and after the successful deployment from Columbia, we required the successful operation of the IUS. Although this system had an excellent track record, the previous use of an IUS had failed, albeit for reasons that we understood and were confident would not reoccur. Still, one worried. Obviously the IUS successfully fired and placed the Observatory near its final orbit. Then followed the use of Chandra's own propulsion system, fired five times over many days to achieve the final orbit. Of course, activation and checkout were taking place in parallel. Perhaps the event that was most intense was the opening of the door to ACIS in view of the previously mentioned failure during test. The ACIS team made a number of technical and operational changes to the door mechanisms and provided more robust opening procedures, but we were unsure that we really understood the root cause of the previous failure. One can imagine that we were very nervous during the opening process. Figure 13 shows a partial view of the control center during the commissioning: One can see the concentration during these times.

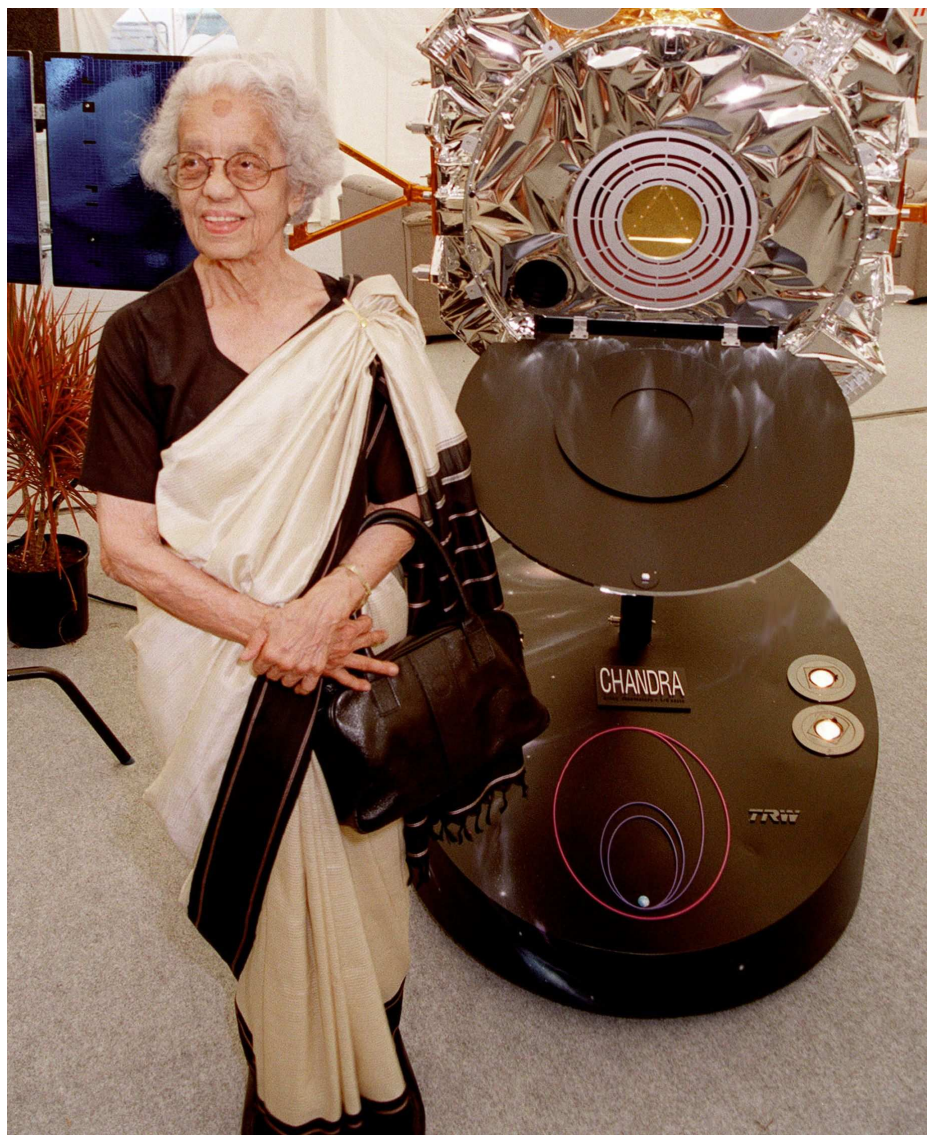

**Fig. 12.** Mrs. Chandrasekhar with a model of the Observatory.

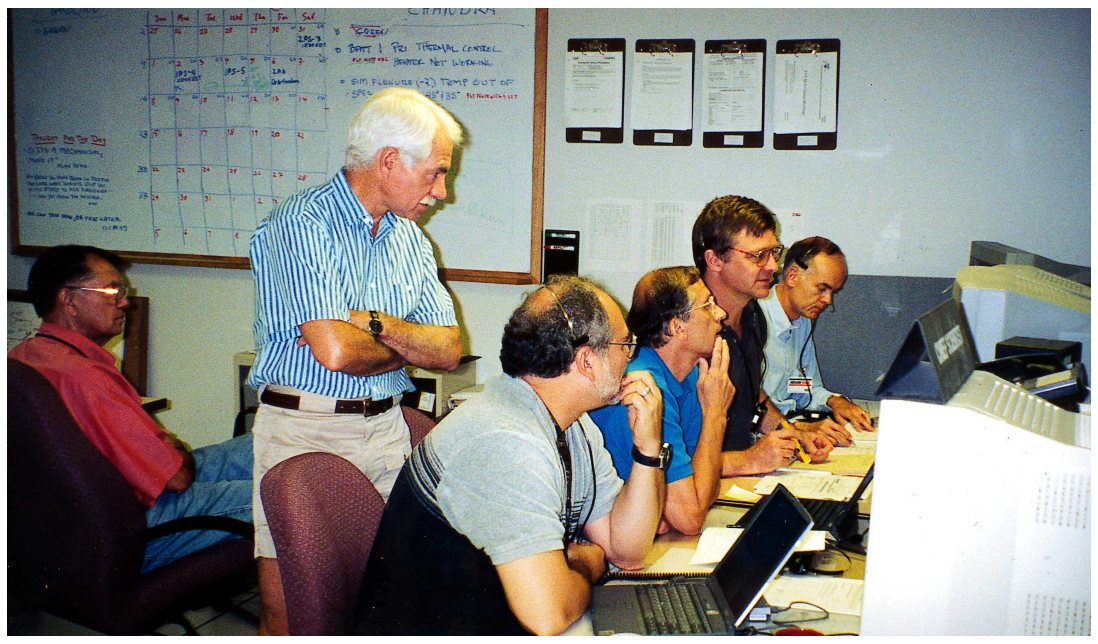

**Fig. 13.** At the Chandra Control center: Left to right: J. Olivier (Deputy Project Manager), C. Canizares (HETG PI), S. Murray (HRC PI), H. Tananbaum (CXC Director), J. MacDougal (MSFC Chief Engineers Office), and R. Schilling (TRW Deputy Project Manager).

The Observatory was launched with S3, the best of the two ACIS back-illuminated CCDs, at the prime focus. Which instrument to place at the focus during launch was the result of much discussion. We wanted to assure a powerful scientific mission in the event that, for some reason, the SIM motions failed thereby preventing changing instruments.

On August 12, 1999 the last door preventing the optics and the ACIS from viewing the universe, was opened. The true "first light" Chandra image is shown in Figure 14. The raw ACIS image of the brightest source on S-3 told us, by inspection — because the flux was concentrated into a handful of AXIS pixels each subtending 0.5 x 0.5 arcseconds — that the Observatory was operating more or less as expected, even prior to establishing best focus. Of course, we now know that the Observatory is performing as predicted. The full S3 field is shown in Figure 15. The bright source which we dubbed "Leon X-1" turned out to be a type-1 AGN at a redshift of z=0.3207. The ubiquity of AGN in X-ray images was not a surprise, but certainly a forerunner of the fabulous discoveries that Chandra was about to make.

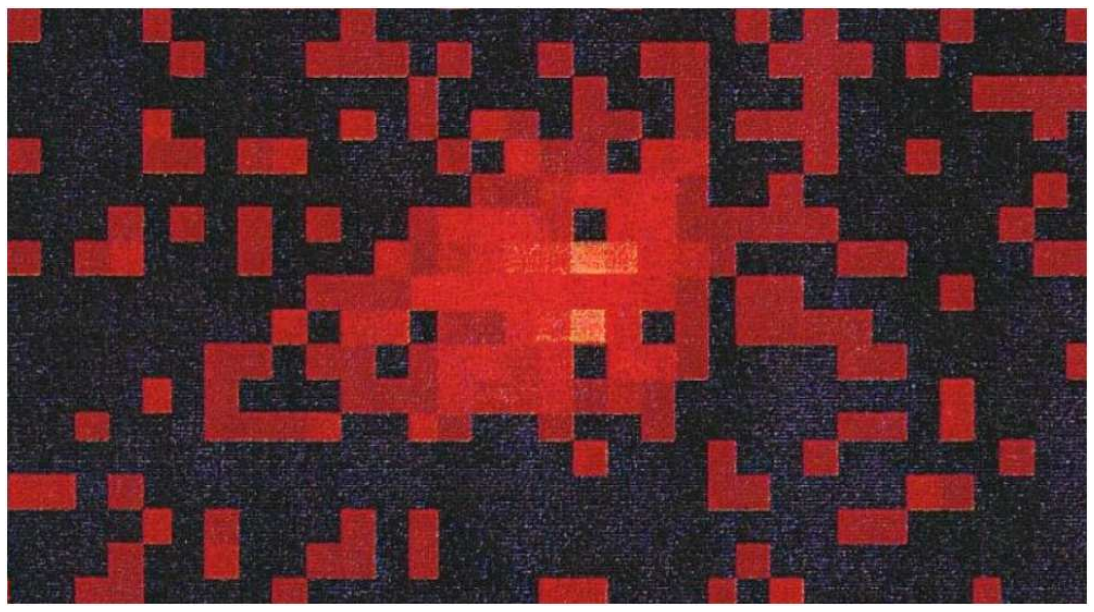

**Fig. 14.** Raw image of "Leon X-1". The data were not yet corrected for spacecraft motion nor was the detector yet at best focus.

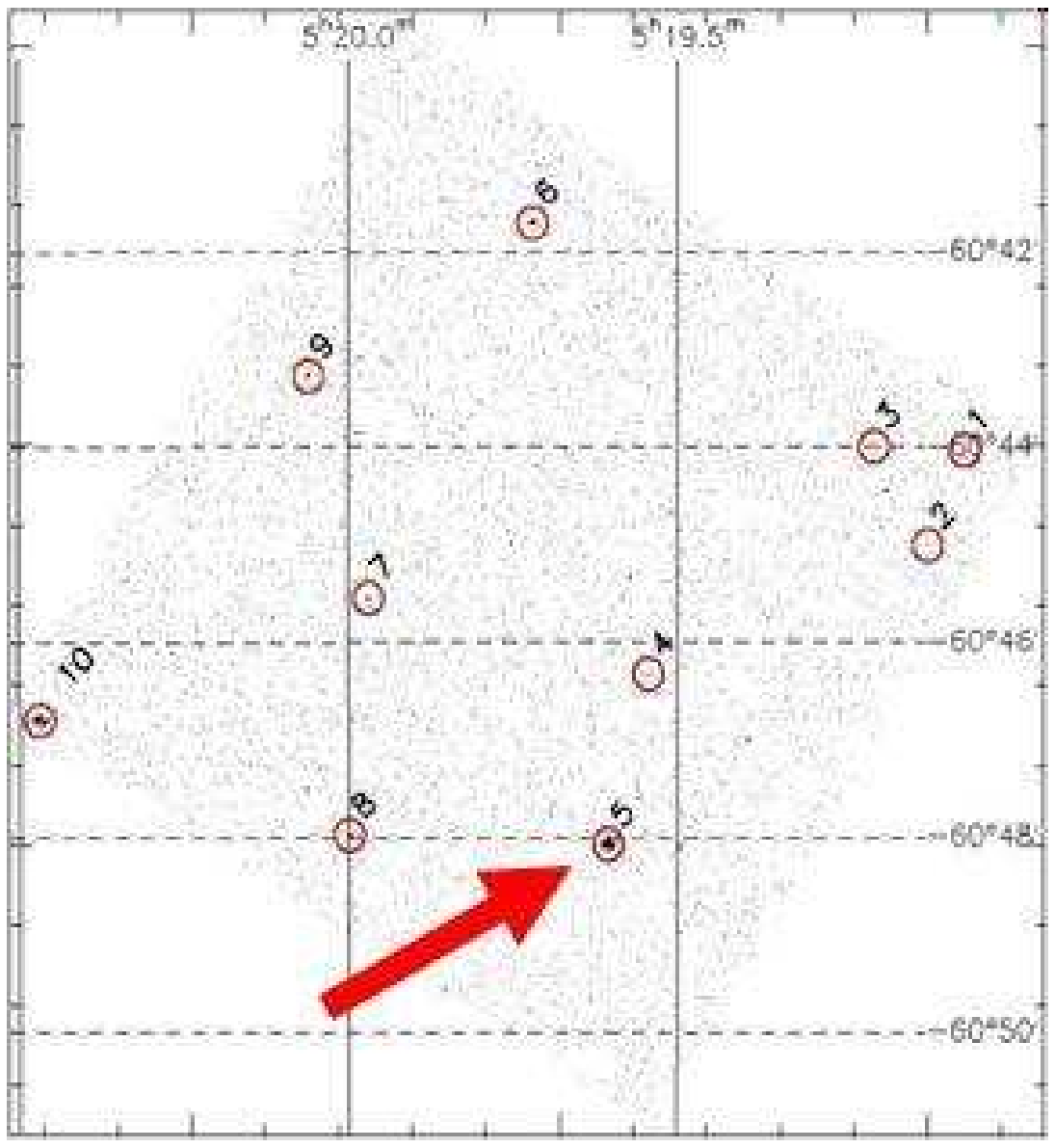

**Fig. 15.** The full "first-light" image on ACIS-S3 after aspect correction. The bright source indicated by the red arrow is "Leon X-1", a type 1 AGN at z=0.3207.

The next tasks in the commissioning process were to test the aspect system and to determine the best focus for ACIS-S. Our target was a bright AGN, selected because we wanted a bright point source. Figure 16 shows the image and highlights the discovery of an X-ray jet, an accidental but certainly exciting outcome. Moreover, the angular resolution of Chandra allowed the use of these data to determine the best focus. Another Chandra "test image" produced a fruitful scientific result. The now famous image of the Crab Nebula, its pulsar, and the remarkable structure showing the shock produced by the pulsar wind was another spectacular result from such a test.

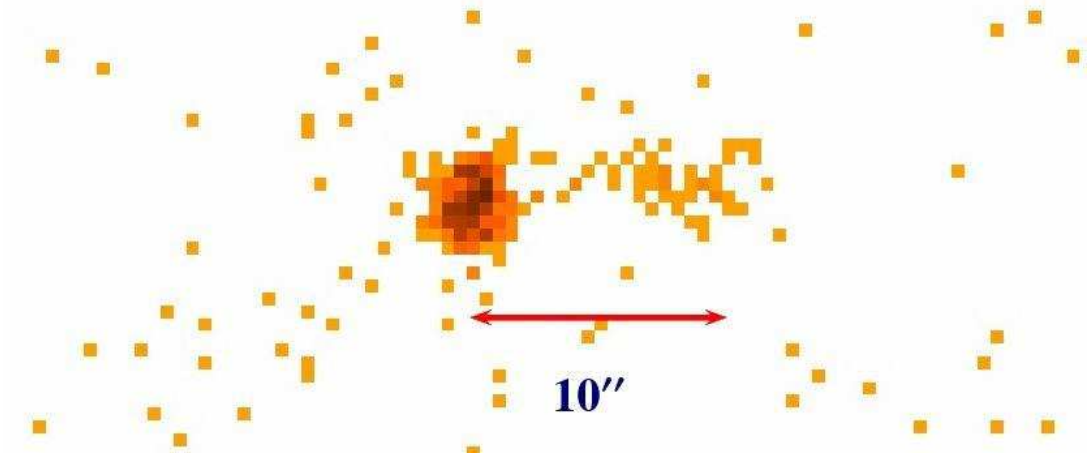

**Fig. 16.** Image of Parkes 0637-752 and its X-ray jet.

Now we are celebrating 10 years of outstandingly successful scientific research with this Great Observatory. There can be no question as to its success, whether measured in terms as mundane as publications, citations, PhD theses, etc., or more profoundly by producing such results as the clarification of the mechanism producing the X-ray emission from comets, to the resolution of the diffuse background, to the independent and confirmatory measures of the Hubble constant and constraining the equation of state of the Universe. In 47 years we have been privileged to participate in the advancement of a discipline that has moved from discovery of the first extra-solar X-ray source to the detection of sources approximately 10 orders of magnitude fainter. The smiles of those pictured below (Figure 17) are as applicable today as they were at the time one saw the first Chandra images.

Those interested in more information on making the Chandra X-ray Observatory should see Wallace and Karen Tucker's book "*Revealing the Universe: The Making of the Chandra X-ray Observatory*", Harvard University Press, 2001.

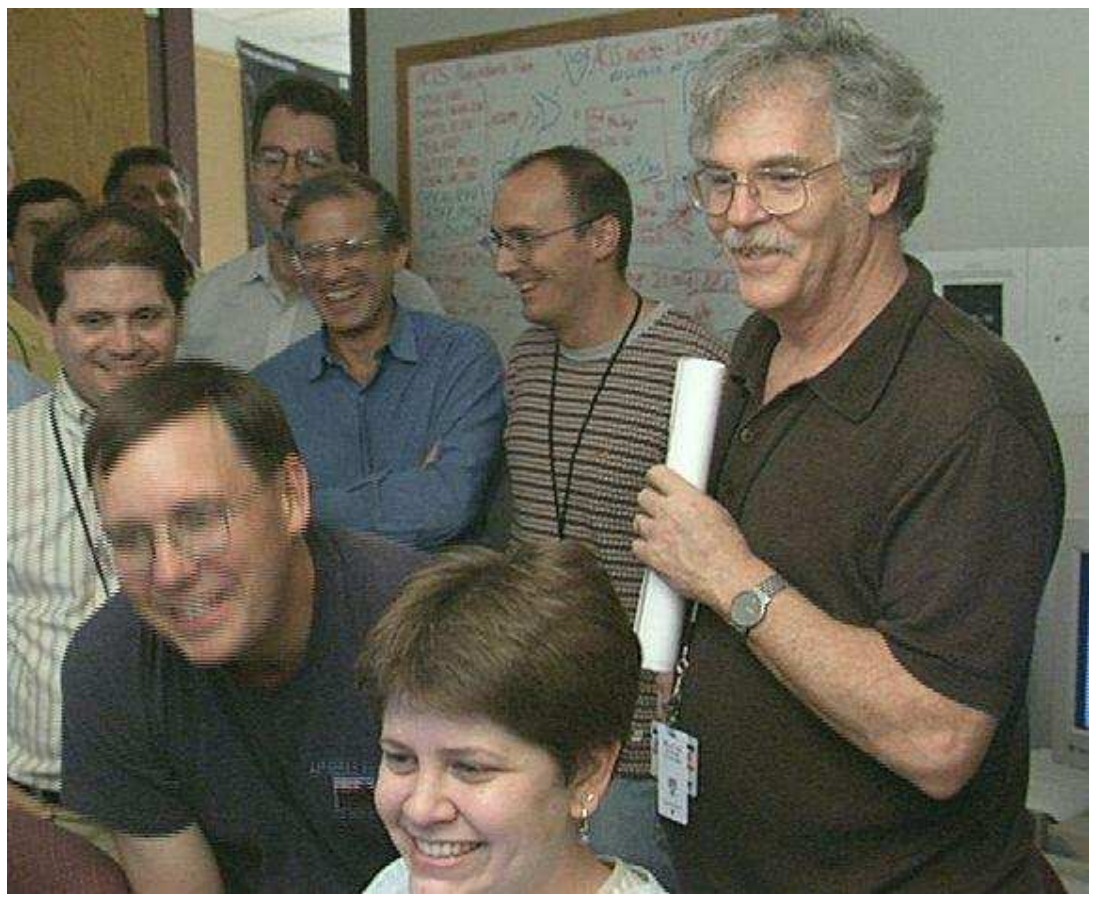

**Fig. 17.** Observing the official first light on August 19, 1999 at the control center. From right to left: M.C. Weisskopf, T. Aldcroft, C. Grant, H. Tananbaum, R. Brissenden, M. Bautz, M. Freeman, F, Baganoff, and K. Gage.

**ACKNOWLEDGMENTS.** We all owe a debt of gratitude to the many people who contributed to Chandra's success including the US taxpayers, who, through their Government, have had the courage and insight to foster such important research.